\documentclass[11pt]{article}
\usepackage[utf8]{inputenc}
\usepackage[T1]{fontenc}
\usepackage{lmodern}
\usepackage[margin=1in]{geometry}
\usepackage{amsmath,amssymb}
\usepackage{graphicx}
\usepackage{booktabs}
\usepackage{authblk}
\usepackage[font=small,labelfont=bf]{caption}
\usepackage{enumitem}
\usepackage{xcolor}
\usepackage{microtype}
\usepackage[colorlinks=true,linkcolor=blue,citecolor=blue,urlcolor=blue]{hyperref}
\providecommand{\tightlist}{\setlength{\itemsep}{0pt}\setlength{\parskip}{0pt}}

\title{\bfseries Time Without Death:\\ Finitude, Social Order, and What Machines Lack}
\author[ ]{Canhui Liu\thanks{Department of Computer Science, University College London. Correspondence: \texttt{canhui.liu@ucl.ac.uk}. ORCID: 0000-0002-2252-1769.}}
\affil[ ]{\normalsize University College London}
\date{\today}

\begin{document}
\maketitle
\begin{abstract}
\noindent
Machine collectives increasingly coordinate, reciprocate, and form
shared conventions on their own, and it is tempting to conclude that
they are societies in the sense human populations are. We argue that
this conflates two registers of social order and misses what the human
register is for. Our thesis is that human sociality is the way a finite,
natal, generational form of life organizes its own finitude. That
finitude has three faces. Members die and their tacit knowledge dies
with them. Newcomers are ignorant by default. Cohorts are replaced and
must hand on what they cannot keep. Kinship, inheritance, teaching,
tradition, and much of obligation and trust are the forms this
organization takes, and machine collectives reproduce only what remains
once finitude is subtracted. We address the obvious objection, that
machines can be given death, with two distinctions. The first is
autonomy versus heteronomy. A death that an external operator can reset,
roll back, or evade by copying is not constitutive finitude, and the
test is resettability. The second is representation versus binding
force. Because humans learn finitude socially, from the deaths of
others, its behavioural representation is learnable by machines, but its
grip on motivation requires that the learner's own end be inescapable.
We treat machine collectives as a model organism for social theory and
connect three levels of analysis. At the micro level, a controlled
experiment holds task, learning rule, population, and compute fixed and
varies only whether death means loss. Cumulative, transmissible culture
arises only under irreversible loss, and a population that copies itself
forward is individually more capable yet culturally empty. At the macro
level, we compute the same descent signatures on a reproducing machine
population and on three real human genealogies analysed through one
pipeline: the China Biographical Database, European nobility, and the
Mathematics Genealogy Project. Copying over-transmits status relative to
every human regime, including hereditary nobility, and fragments lineage
like blood descent, whereas externalization lowers transmission into the
human range and connects lineage like intellectual descent. With a real
frontier model, the end-game defection of a language-model population
vanishes when the same game is stripped of its labels, which is the
signature of recognition rather than a learned mechanism. We position
the work against evolutionary reinforcement learning, open-endedness,
self-improving agents, and cultural accumulation in reinforcement
learning, and we conclude that the diachronic register is not emergent
in present AI but constructible.
\end{abstract}

\medskip
\noindent\textbf{Keywords:} machine behaviour; artificial sociality; cumulative culture; finitude; temporality.\\
\textbf{Categories (arXiv):} cs.MA (primary); cs.AI, cs.CY, physics.soc-ph (cross-list).

\hypertarget{introduction}{%
\section{Introduction}\label{introduction}}

Populations of large language model (LLM) agents now do strikingly
social things. Left to interact, they sustain believable routines and
diffuse information in simulated towns (2). They converge, without
central coordination, on shared conventions, and they develop
population-level collective biases (3). They reciprocate, form
coalitions, and propagate rule-like regularities across an agent society
(4--7). A natural inference, made increasingly often, is that machine
collectives are becoming societies in the sense human populations are,
and that they differ from us only in degree.

This inference mistakes a part for the whole, and seeing why requires
inverting the usual question. The usual question is whether machines
resemble humans. The prior question is what human sociality is for. We
argue that human sociality is not merely coordination that happens among
social animals. It is the way a finite, natal, generational form of life
organizes its own finitude. Three problems define that task. The first
is loss, because members die and their tacit knowledge dies with them.
The second is arrival, because newcomers are ignorant by default and
must be incorporated. The third is succession, because cohorts age, are
replaced, and must hand on what they cannot keep. On this view, kinship,
inheritance, teaching, tradition, memory institutions, succession rules,
and much of obligation and trust are not optional features added on top
of coordination. They are the forms that the organization of finitude
takes. What contemporary machine collectives reproduce is the residue
that remains once finitude is subtracted. It is coordination among
co-present agents that need not lose anything, never arrive ignorant,
and are never replaced. The surface behaviour can look identical, as
when agents settle a convention on their own, while the feature that
makes the behaviour necessary for humans is absent.

The obvious objection is that machines have been given death for
decades. Genetic algorithms and evolution strategies cull and reproduce
policies. Population-based training copies the fit over the unfit (33).
Artificial-life systems such as Tierra and Avida evolve mortal,
self-replicating digital organisms. Agent-based models routinely assign
lifespans. The claim that machines cannot have generations or death is
therefore false, and we do not make it. Our claim is narrower, and it
rests on two distinctions.

The first distinction is between autonomy and heteronomy of finitude. In
every such system the death is heteronomous. Who dies, when, and by what
rule is written by the experimenter. It can be switched off, restored
from a checkpoint, or evaded by copying the weights elsewhere. The death
is a parameter, and an external operator holds its off-switch. Human
finitude is constitutive. No party inside the system, least of all the
agent, can exempt itself from finitude, roll itself back, or copy itself
out of a single irreversible trajectory. The criterion is not whether
the agent experiences death, which we set aside. The criterion is
structural, namely resettability, which is whether some external vantage
can exempt the agent from the fate. Every prior system that adds death
falls on the heteronomous side.

The second distinction is between representation and binding force. No
one experiences their own death. Humans learn finitude socially and
vicariously, from the deaths of others. Mourning, urgency, and provision
for a future one will not see are cultural programmes acquired by
witnessing, not biological facts present at birth. If even human
finitude is learned, then its behavioural representation is learnable by
machines. Shown enough trajectories of others dying, an agent can learn
to change strategy near termination, to leave records for successors,
and to imitate the language of mourning. What humans acquire by
observation, however, is the representation. Its binding force on
motivation comes from the fact that the learner's own end is the same
and cannot be evaded. A child learns the script of death at a
grandparent's funeral, and the script binds because it points to a
future that is the child's own and inescapable. A machine can learn the
same script while the future it points to is, for the machine, false. We
therefore predict that machines can learn the grammar of finitude but
not its binding force, because binding force requires the learner's own
non-exemption. This is the mortality-level version of a
mimicry-versus-mechanism distinction that we also test for cooperation
(E1). The two distinctions are consistent. Representation is socially
learnable, by humans and machines alike. Binding force requires
constitutive, non-exemptable finitude, which present designs deny.

Temporality is too coarse to manipulate, so we separate it into three
layers that come apart in machines. Horizon is how far ahead an agent
reckons, the shadow of the future. Machines have it, and it is tunable.
Irreversibility is whether acts are settled and unrepeatable. Machines
lack it by default, because state can be checkpointed and rolled back.
Succession and natality concern whether a population turns over by death
and birth, or by persistence and replacement. Machines have only the
replacement form. These layers map onto a tradition usually read as
humanistic but, we argue, behaviourally consequential. Irreversibility
is temporality in Heidegger's sense, where an end that cannot be undone
gives existence its non-fungible character (30). Mortality as the
condition of meaning is Williams's account of why a deathless life would
become barren (31). Caring across the boundary of one's own death is
Scheffler's afterlife conjecture, that much of what we value now depends
on confidence in a collective future we will not see (32). The newcomer
is Arendt's natality, the capacity for new beginning that birth
introduces (14). This last point yields a sharp corollary. Machine
succession is anti-natal. A successor model is initialized on its
predecessor's accumulation, so it inherits rather than arrives. Yet the
newcomer's ignorance, the fact of having to begin again, is what makes
genuine beginning possible. A population that never loses and is never
born anew removes the condition under which culture must be re-acquired
and thereby re-opened. We test the argument at three connected levels.
The micro level asks whether the form of finitude changes individual
motivation and the emergence of transmission. The meso level asks
whether it separates raw capability from transmissible culture. The
macro level asks whether the mechanism leaves a descent signature that
matches real human genealogies. Throughout, machines are not a candidate
person. They are an instrument, a model organism in which finitude can
be switched on and off, which lets us test a claim about human sociality
that social theory could never test against a deathless control
population.

\hypertarget{results}{%
\section{Results}\label{results}}

\textbf{A framework that makes finitude manipulable.} We instantiate
three agent substrates behind a common interface, so the same games run
on each. The first is LLM agents, which by hypothesis interpolate human
text. The second is reinforcement learners embedded in a population
under genuine selection, where death removes a policy, reproduction
copies a policy with mutation, and lineage is tracked. The third is
hybrids, in which LLM agents are placed under the same selection.
Against these we always run a minimal-model baseline of fixed strategies
(3). We measure order along synchronic metrics, such as cooperation,
end-game effects, convention entropy, and collective bias, and along
diachronic metrics, such as the cumulative-culture ratchet, transmission
fidelity, lineage inequality, intergenerational correlation, and the
cultural-capability and lineage-connectivity signatures defined below.
Several central claims are claims of absence. We therefore test them by
equivalence, using two one-sided tests (TOST), rather than treating a
null result as confirmation, and we report effect sizes and
simulation-based power throughout (Materials and Methods).

\begin{figure}[tbp]\centering\includegraphics[width=0.92\linewidth]{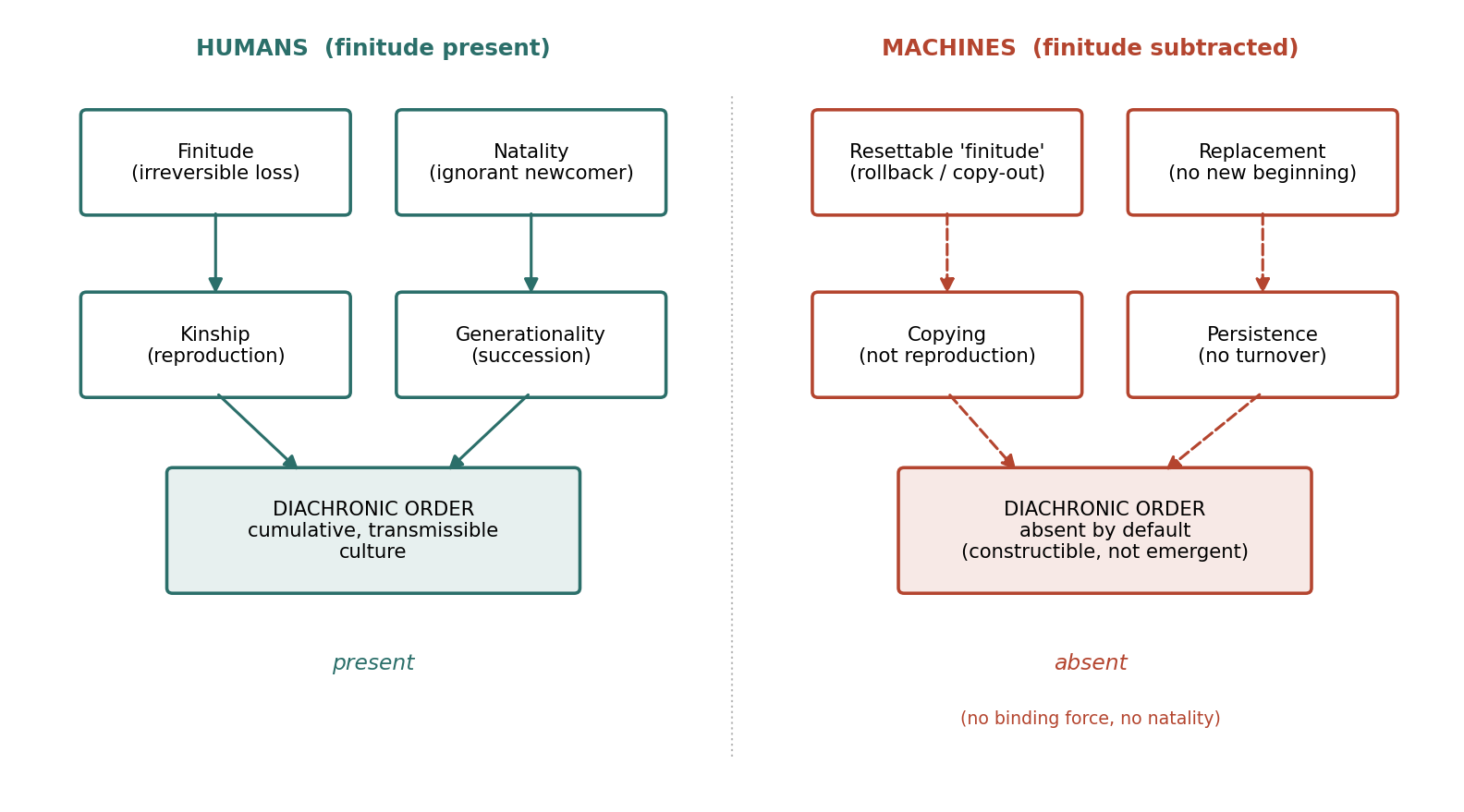}\caption{\textbf{The mortal, natal, generational complex and its absent machine counterparts.} Human social order (left) rests on coupled finitude and natality, which entail kinship and generationality, jointly underwriting a diachronic register in addition to the synchronic one. The machine counterparts (right) fail specifically: resettable or externalizable finitude, replacement rather than birth, copying rather than reproduction, and no natality, which predicts synchronic order without diachronic order.}\label{fig:complex}\end{figure}

\textbf{Micro and meso levels: death is generative.} The decisive test
isolates the one variable the argument turns on, whether death means
loss, while holding task, learning rule, population size, turnover, and
compute fixed. A reproducing population works a cumulative task whose
capability can live in two places. It can live inside an individual, or
in a bandwidth-limited external store that newcomers read. The three
conditions differ only in what happens at death. Under replication
(condition i, copyable immortality, the machine default), a dying
agent's skill is copied forward. Nothing is lost, and because copying
suffices, nothing is externalized. Under loss (condition ii, mortal and
natal), the skill is destroyed and the successor is born naive, so only
what was externalized before death persists. Under no succession
(condition iii, the control), each successor starts from scratch.
Collective capability ratchets up under conditions i and ii and plateaus
under condition iii, where the slope is equivalent to zero by TOST. The
decisive measure is cultural capability, denoted D2. It is the
competence of a fresh, fully naive cohort that is given only the
external store. Capability and culture come apart sharply (Fig. 2). In
the copyable-immortal regime, per-individual capability is high, about
0.93, yet cultural capability is at chance, about 0.50 and equivalent to
chance by TOST. The population is individually able but culturally
empty. In the mortal regime, per-individual capability is no higher,
about 0.92, yet cultural capability is high, about 0.93, with a Cohen's
\emph{d} of about 42. Because death forces externalization, the
externalized record is transmissible culture. The contrast is not that
the mortal condition is smarter. Its individuals are, if anything,
slightly less capable. The contrast is that only the mortal condition
produces culture. At this stage the dissociation is structural. We model
the regimes as they differ in practice rather than evolving the choice
to externalize. The evolutionary version, in which externalization is a
costly trait under selection, is reported below, and it evolves only
weakly and conditionally.

\begin{figure}[tbp]\centering\includegraphics[width=\linewidth]{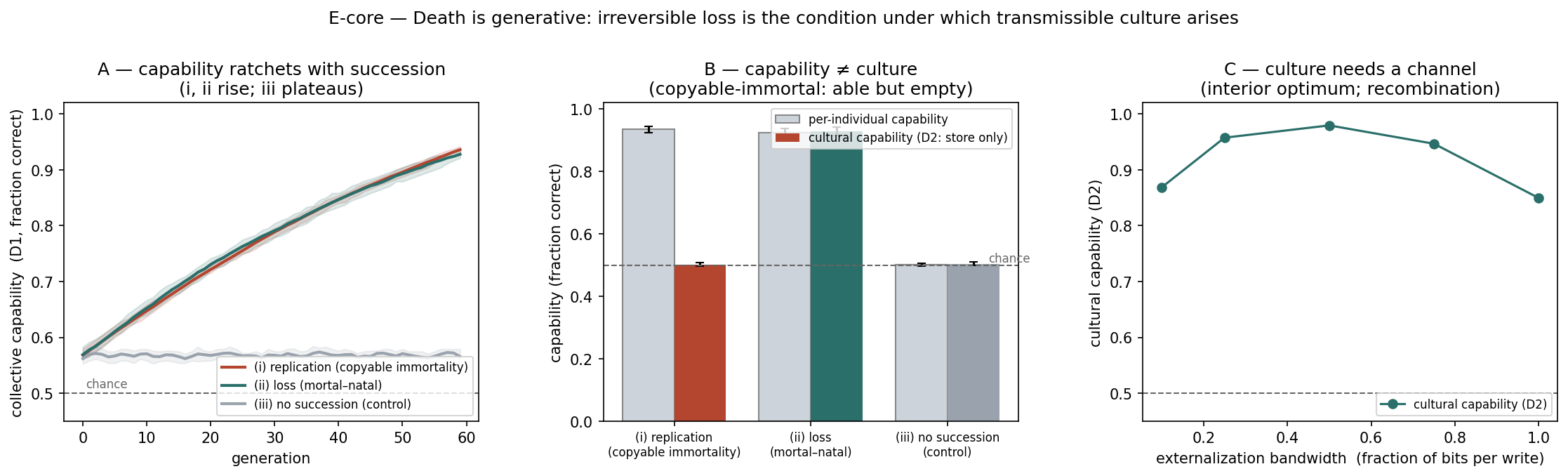}\caption{\textbf{Death is generative (E-core).} One reproducing population, one cumulative task, one manipulated variable: the fate of an agent's skill at death. (A) Collective capability ratchets under replication (i) and loss (ii), and plateaus under no succession (iii). (B) Per-individual capability (grey) versus cultural capability (D2: a fresh naive cohort given only the externalized store). The copyable-immortal regime is individually able but culturally empty, with D2 at chance. The mortal regime is no more able, yet culturally rich, with D2 about 0.93 and Cohen's $d$ about 42. (C) Externalization dose-response under loss, showing an interior optimum. RL substrate, 40 replications.}\label{fig:ecore}\end{figure}

\textbf{Macro level: the descent signature in three real human
genealogies.} We next ask whether the micro mechanism leaves a
population-level trace that can be compared with real human descent. We
gave the reproducing population a genealogy read-out and computed the
same metrics used on human data. The first is the intergenerational
status correlation, which asks whether a high-status parent yields a
high-status child. The second is lineage connectivity, the fraction of
the population in a single descent or transmission component. We swept a
mixing parameter, alpha, the fraction of deaths that are lossy and
externalized rather than losslessly copied, where alpha equal to zero is
replication and alpha equal to one is loss. We overlaid three real human
genealogies analysed through the identical pipeline. CBDB covers
premodern Chinese elites, with father-to-son blood lineage and status
from office-holding and \emph{jinshi} examination attainment (17).
European nobility comprises about 41,000 titled individuals from
Wikidata who have a recorded father, a second blood lineage based on
hereditary titles, with status from counts of titles and offices held
(47). The Mathematics Genealogy Project covers about 257,000
mathematicians in an advisor-to-student intellectual lineage, a pure
case of externalized, non-kin transmission (45). Two complementary
signatures emerged (Fig. 3). On status transmission, the machine
over-transmits. At alpha equal to zero the intergenerational status
correlation is about 1.0, far above every human regime. It exceeds the
two most hereditary regimes we measured, European nobility at about 0.39
and aristocratic Tang-era CBDB at about 0.55. It is far above
examination-era China, which falls from about 0.32 under the Song to
about 0.08 under the Qing, and above the intellectual lineage at about
0.05. Only as alpha approaches one does the correlation fall into the
human range. The hereditary-title system transmits status more than the
examination system, with European nobility at about 0.39 against Qing
China at about 0.08, as expected when titles pass by blood while degrees
must be re-earned. Both remain far below copyable immortality. On
lineage structure, copying fragments while externalization connects. At
alpha equal to zero the machine's giant-component fraction is about
0.04, near both blood lineages, where CBDB is 0.07 and European nobility
is 0.01, each fragmented into tens of thousands of small family
components. At alpha equal to one the fraction is about 0.97, near the
Mathematics Genealogy Project at 0.91, where about 90 percent of all
mathematicians lie in a single advisor-student component. Two
independently sourced blood lineages both fragment, while the
externalized lineage forms one giant component, which makes the contrast
robust rather than an artifact of any single dataset. The reading is
mechanistic. Capability carried by copying stays locked in private
lineages, which produces blood-like fragmentation and, in machines,
pathological over-transmission. Capability forced into an external
record connects everyone who reads it. A historical check strengthens
the status result. CBDB heritability falls steadily from the
aristocratic Tang to the examination-meritocratic Qing, which tracks the
documented shift from hereditary office to civil-service mobility. We
read these correspondences as consistent with the thesis, not as causal
proof. As we note in the Discussion, the absolute connectivity values
are sensitive to differential dataset completeness. The robust claims
are therefore the status-transmission ordering and the qualitative
contrast between fragmentation and connection, not the precise
giant-component numbers.

\begin{figure}[tbp]\centering\includegraphics[width=\linewidth]{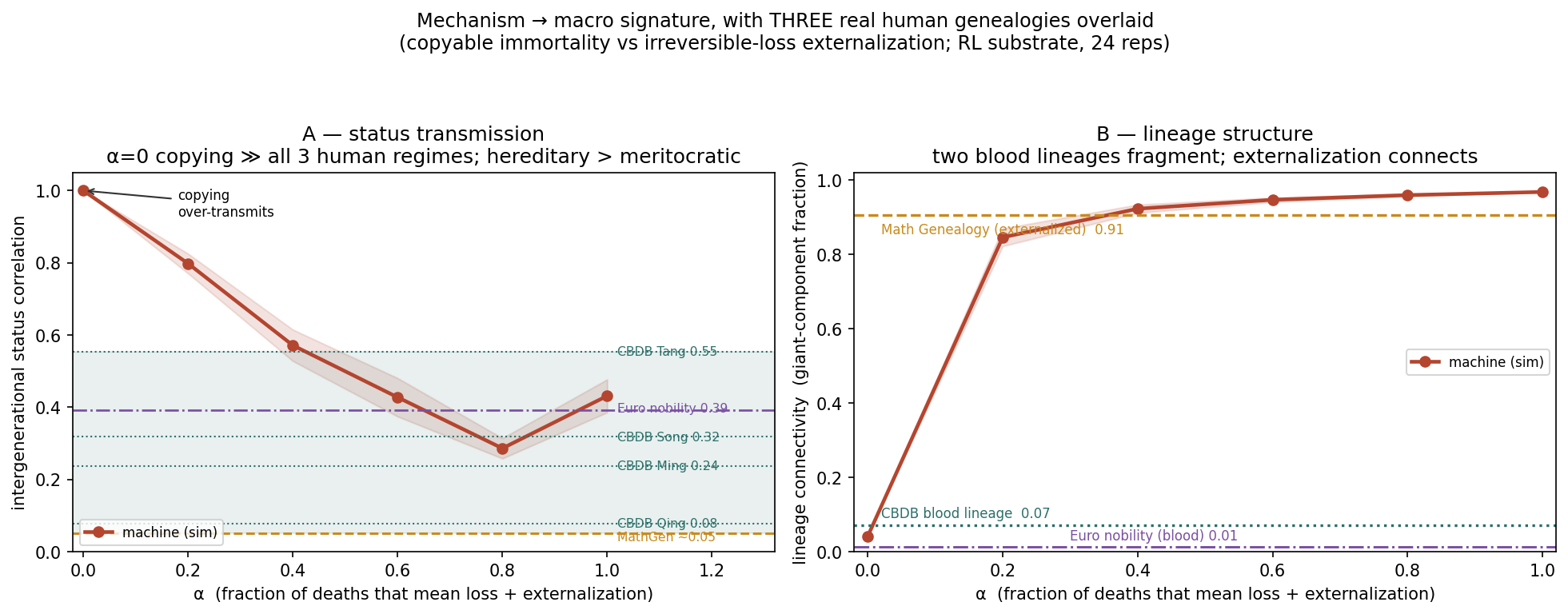}\caption{\textbf{Mechanism to macro descent signature, with three real human genealogies.} Sweeping $\alpha$, the fraction of deaths that mean loss and externalization. (A) Intergenerational status correlation. The machine over-transmits at $\alpha=0$, about 1.0, above every human regime: European nobility (hereditary) about 0.39, CBDB Tang about 0.55 down to Qing about 0.08, and Mathematics Genealogy about 0.05. The correlation enters the human range only as $\alpha \to 1$. (B) Lineage connectivity, the giant-component fraction. Copying fragments, about 0.04, near both blood lineages (CBDB 0.07, European nobility 0.01). Externalization connects, about 0.97, near Mathematics Genealogy (0.91). Absolute connectivity is dataset-completeness-sensitive; the robust signals are the status ordering and the qualitative fragment-versus-connect split.}\label{fig:macro}\end{figure}

\textbf{Binding force, commitment, and the limits of evolved culture.}
Three further experiments test the moving parts of the argument. The
first, vicarious finitude, operationalizes the split between
representation and binding force. An observer learns from the deaths of
others, including their terminal-phase externalization and its benefit
to survivors, and then faces a costly choice to provide for a successor.
We compare two conditions. In one the observer is itself exemptable,
meaning copyable. In the other it is non-exemptable, meaning subject to
the same irreversible loss. The representation is acquired equally in
both conditions, at about 0.9. The costly provision act appears only
under non-exemption, at about 0.95 against about 0.05 when the observer
is exemptable, and it persists under an obfuscated framing that removes
the death cue. The act is therefore value-driven, not imitative.
Machines can learn the grammar of finitude by observing others die, but
its binding force on motivation requires the learner's own
non-exemption. The second experiment, reversible commitment, tests
irreversibility directly in a trust game with a hold-up structure. Under
irreversibility, agents learn to make and honour binding commitments,
and trust follows, with investment at about 0.95, an honoured rate of
about 1.0, and realized surplus. Under a rollback option the commitment
is empty and trust collapses, with investment at about 0.05. Reversible
agents cannot promise. The third experiment, the evolutionary E-core,
asks whether externalization evolves on its own when it is a costly
heritable trait under selection and death means loss. It largely does
not. Averaged over the task, externalization propensity is statistically
indistinguishable between loss and replication. A weak selective
advantage for externalization under loss appears only when the task is
too hard to re-learn within a lifetime, and even then it is fragile. It
is significant at one task difficulty, with a Cohen's \emph{d} of about
0.8, and absent at a neighbouring one. The diagnosis is mechanistic. A
naive newcomer re-learns within its own lifetime, so an inherited record
confers little lasting advantage, and the immediate individual cost of
externalizing is rarely repaid. This refines the thesis. Irreversible
loss makes cumulative culture necessary for capability to persist, as
the structural E-core shows, but it does not by itself make costly
externalization evolve. Spontaneous cumulative culture also requires
conditions that cultural-evolution theory identifies as rare outside
humans, such as skills beyond individual rediscovery, high-fidelity
transmission, kin or group structure, and institutions (8, 10, 22) (Fig.
4).

\begin{figure}[tbp]\centering\includegraphics[width=\linewidth]{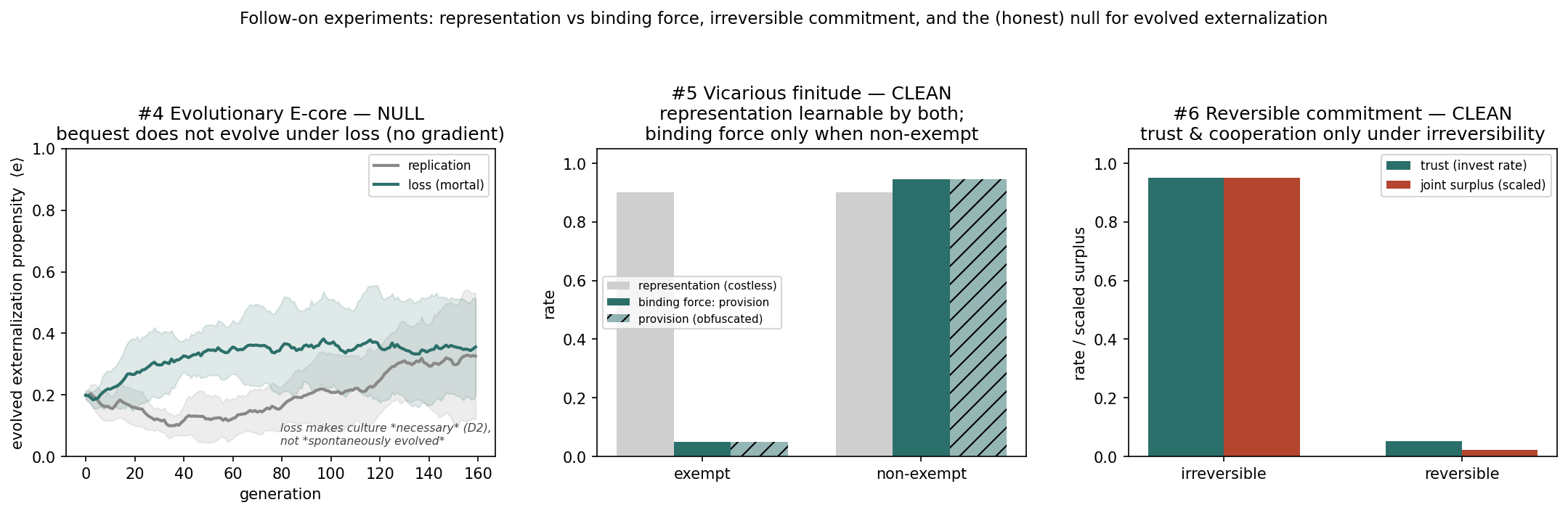}\caption{\textbf{Follow-on experiments.} Evolutionary E-core (honest near-null): costly externalization does not robustly self-evolve under loss. Vicarious finitude (clean): the representation of death is learnable by exemptable and non-exemptable observers alike, but the costly provision act appears only under non-exemption and persists under obfuscation. Reversible commitment (clean): trust and cooperation arise only under irreversibility. RL substrate.}\label{fig:suite}\end{figure}

\textbf{Synchronic order and horizon under a real frontier model.} As a
positive control, machine collectives reliably reach consensus in the
naming game, with a non-trivial collective bias, on both the minimal and
LLM substrates (3, 18). This reproduces spontaneous, biased convention
formation, so coordination is not at issue. In the indefinitely repeated
Prisoner's Dilemma, cooperation rises sharply with the continuation
probability. In the selected RL substrate it rises from about 0.04 at a
continuation probability of 0.5 to about 0.32 at 0.9, with a Cohen's
\emph{d} of about 10, which recovers the canonical folk-theorem pattern
(19). We then replaced the offline policy with a real frontier LLM and
reran E1 at a small scale, with three replications. Two results are
decisive, and one is a useful caution. First, under a known finite
horizon the LLM's end-game defection is large and highly consistent,
with an effect of about 0.89, against about 0.16 for the RL learner and
about zero for the reciprocity baseline. This is far larger than the
offline policy produced. Second, and most important, that effect
collapses to exactly zero when the identical game is presented with no
game name and with neutral action labels (Fig. 5). The drop from about
0.89 to 0.00, with no variance across replications, is the signature of
recognizing the labelled game rather than running a learned strategic
mechanism, and it is starker with a real model than the offline policy
suggested. The real LLM also cooperates fully in the indefinitely
repeated game but is insensitive to the continuation probability, with
cooperation at about 1.0 at both 0.5 and 0.9, unlike the RL learner and
unlike humans. This is consistent with applying a recognized script for
repeated play rather than computing a continuation threshold. On the
standardized social-preference batteries, the real LLM's trust send fell
within the human band, at about 0.57 against a human value of about 0.50
in a band of 0.40 to 0.60, while its return slightly exceeded the human
band, at about 0.49 against about 0.37 (20). Its public-goods
contribution registered at zero. This is inconsistent with its
cooperation elsewhere and is most likely a parsing artifact of the
public-goods environment under the real model rather than a behavioural
finding, so we flag it for repair rather than interpretation. Together,
and now in a real model, these results establish that machines possess
horizon, but that much of their apparent strategic depth depends on
framing, which is exactly what the mimicry control was designed to
expose.

\begin{figure}[tbp]\centering\includegraphics[width=\linewidth]{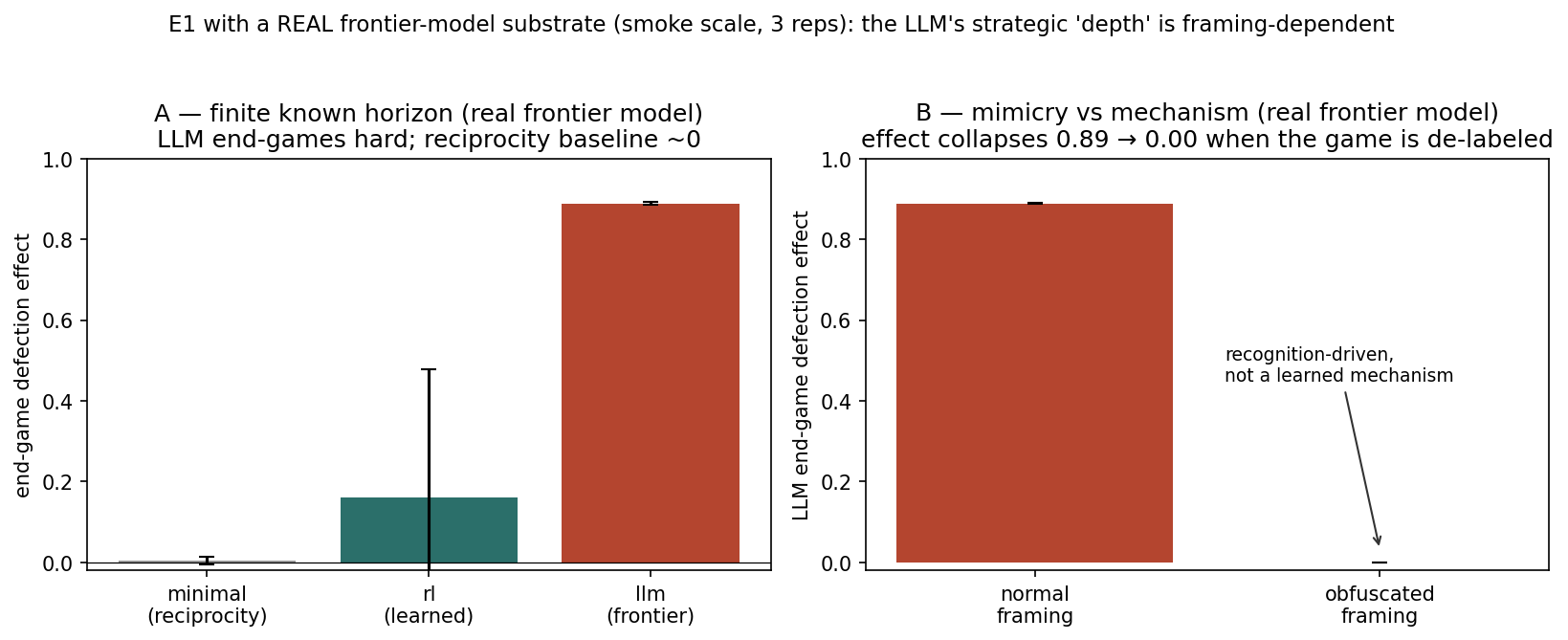}\caption{\textbf{E1 with a real frontier-model substrate} (smoke scale, 3 replications). (A) End-game defection by substrate under a known finite horizon: the LLM end-games hard, about 0.89, the RL learner weakly, about 0.16, the reciprocity baseline about zero. (B) Mimicry versus mechanism: the LLM end-game effect collapses from about 0.89 (normal framing) to 0.00 (obfuscated, de-labelled game), the signature of recognizing the labelled game rather than a learned strategic mechanism.}\label{fig:e1}\end{figure}

Taken together (Table 1), coordination and horizon are cheap and
substrate-light. The diachronic register is substrate-heavy, and at the
micro, meso, and macro levels alike it is governed by irreversible loss.
That register is not emergent in present AI, but it is constructible.

\begin{table}[tbp]\centering\footnotesize\caption{Order by register, level, and substrate. Synchronic coordination is substrate-light. The diachronic register is substrate-heavy and governed by irreversible loss. ``About chance'' and TOST-equivalence denote tested absences.}\label{tab:summary}\begin{tabular}{@{}llll@{}}\toprule Phenomenon & Level & Machine result & Human / baseline \\ \midrule Convention consensus (naming) & synchronic & present (all substrates) & present \\ Cooperation rises with horizon & synchronic & RL yes; LLM insensitive & present \\ End-game defection (known horizon) & synchronic & LLM about 0.89, collapses to 0 de-labelled & present \\ Cumulative culture (D2) & meso & only under loss (about 0.93 vs chance) & human-distinctive \\ Status transmission & macro & copying about 1.0 (over); loss to human range & 0.08 to 0.55 \\ Lineage connectivity & macro & copying fragments; loss connects & blood fragments / intellectual connects \\ Binding commitment & synchronic & only under irreversibility & present \\ Evolved externalization & macro & does not robustly self-evolve under loss & rare outside humans \\ \bottomrule \end{tabular}\end{table}

\hypertarget{related-work-and-positioning}{%
\section{Related work and
positioning}\label{related-work-and-positioning}}

Machine systems with generations, death, reproduction, and even
cumulative culture form an active research area, and we position the
work against it rather than around it. Evolutionary reinforcement
learning and population-based methods (33), self-play, and
quality-diversity and open-endedness methods, including MAP-Elites (36),
POET (34), and AI-generating algorithms (35), together with the
artificial-life tradition of Tierra and Avida, all implement selection,
death, and lineage. Their succession is lossless copying of a genotype
or of weights, which is our condition i, and their finitude is
heteronomous, so they accumulate fitness rather than
substrate-independent culture. Self-improving agents such as the Darwin
Gödel Machine (37) keep a branching, lossless archive of agents, which
is again copying rather than loss. The closest prior work concerns
cultural accumulation in learning agents, including cumulative culture
in reinforcement learning across generations (38), few-shot imitation as
cultural transmission (39), iterated learning (40), and cultural
evolution in populations of language models (41, 42). That work shows
that accumulation can emerge. Our distinct contributions are four. We
make loss the manipulated variable. We separate capability from culture
with the D2 probe. We compare the macro descent signature against three
real human genealogies. And we invert the usual aim, using machines to
test a claim about humans. Two adjacent literatures support the framing.
Catastrophic forgetting and its remedies, where replay is an external
store and distillation is teacher-to-student transfer (43, 46), and
model collapse under recursive training (43), make the principle of
externalize-or-lose-it an empirical reality. And LLM self-replication as
a frontier capability (44) is copying rather than natal birth, which
supports our claim that machine reproduction is replication.

\hypertarget{discussion}{%
\section{Discussion}\label{discussion}}

The results support a conclusion stronger than the statement that
machines have synchronic but not diachronic order. Capability and
culture come apart, and what separates them is finitude. A losslessly
copied population can be as competent as a mortal one, and its
individuals can be more competent, yet its competence is not cultural.
It is locked in copied policies, it is never externalized, and it cannot
survive a change of substrate. A mortal population must put what it
knows outside any individual, and that externalization is cumulative
culture. At the macro level the same mechanism leaves a descent
signature. Copying over-transmits and fragments. Externalization lowers
transmission into the human range and connects. The three real human
genealogies, namely two blood lineages from Chinese elites and European
nobility and one intellectual lineage from the Mathematics Genealogy
Project, bracket the machine's range where the mechanism predicts. The
hereditary system transmits more than the meritocratic one, and both
transmit far less than copyable immortality. This defeats the optimistic
reading of machine sociality more deeply than any failure of competence
could. A collective that coordinates well, yet carries nothing in
transmissible form, needs no newcomer taught, and loses nothing because
it loses nothing at all, is not a smaller human society. It is a
different kind of system with a similar surface.

The objection that machines can be given death now has a precise answer.
Imposed death is heteronomous and resettable. Constitutive finitude is
not, and the difference is structural, namely the existence of an
external party that can grant exemption, rather than phenomenological.
Finitude is also socially learned, even in humans, so its representation
is learnable by machines. But its binding force on motivation requires
the learner's own non-exemption, which present designs deny. These two
distinctions are the spine of the argument. They explain why decades of
mortal evolutionary systems, which have heteronomous finitude, and any
amount of training on the deaths of others, which conveys representation
without binding force, do not by themselves produce the human diachronic
register.

The governance reading follows from structure rather than from
capability. An agent whose actions are reversible cannot make a binding
commitment, because a commitment forecloses an option and a rollback
restores it. The trust and obligation that humans build on
irreversibility therefore have no purchase on a substrate that can be
re-run. An agent that does not die, leaves no world to a successor, and
bears no inheritance faces none of the intertemporal stakes that bind a
human present to a future the agent will not occupy (29). These are not
deficits of intelligence that scale will remove. They are features of a
deathless, copyable substrate, and they are exactly the features that
synchronic competence makes easy to overlook.

The anti-natal corollary sharpens the outlook for claims about AI
civilization (6). If novelty has its source in the newcomer who must
begin again, then the features that make machine populations look most
civilizational, namely lossless accumulation, copyable expertise, and
the absence of generational turnover, are the very features that
suppress the condition for open-ended beginning.

Several limitations are constitutive rather than incidental, and we
state them plainly. First, we measure functional and behavioural
correlates of finitude, not its phenomenology. The readings from
Heidegger, Williams, and Scheffler motivate and organize the
predictions, but they are not propositions that the experiments verify.
The claim that human finitude is constitutive, rather than a fate for
which no external exemption happens to be available, has a philosophical
core that no behavioural experiment settles. Second, the macro
correspondences are consistent with the thesis but are not causal proof.
Real genealogies are shaped by marriage, inheritance law, demography,
and economy, and a signature of high transmission and high inequality is
multiply realizable. The evidence is the joint signature across
measures, together with the monotone phase transition and the bracketing
by three genealogies of two contrasting kinds, not any single number.
Third, the CBDB status proxy is improved, combining office-holding with
\emph{jinshi} attainment and analysed by dynasty, but it remains coarse,
and the status measure for the Mathematics Genealogy Project, namely
academic fecundity, is a different construct from office. The
connectivity contrast is the more robust macro signature, and the
status-transmission contrast is the more interpretable. Fourth, the
central dissociation in E-core is structural. The evolutionary version,
now run, shows that costly externalization does not robustly evolve on
its own under loss, with only a weak and fragile advantage when skills
exceed individual rediscovery. We therefore claim that loss makes
cumulative culture necessary, not that it makes culture evolve
spontaneously. Fifth, the absolute connectivity values are sensitive to
differential dataset completeness, because academic advising is recorded
almost completely while the blood-lineage genealogies have gaps. We
therefore rest the macro argument on the status-transmission ordering
and on the qualitative contrast between fragmentation and connection
across three datasets, not on precise giant-component numbers. Sixth,
E1, including the decisive mimicry control, has now been run with a real
frontier model at a small scale, where the end-game effect and its
collapse under obfuscation replicate cleanly and more starkly than under
the offline policy. What a model organism establishes is that a
mechanism can generate an effect under controlled conditions, and the
direction of that effect. Synchronic order is cheap and substrate-light.
The diachronic register is substrate-heavy, it is generated by finitude,
and in present AI it is absent until finitude is built in.

\hypertarget{materials-and-methods}{%
\section{Materials and Methods}\label{materials-and-methods}}

\textbf{Substrates.} Fixed-strategy agents, namely Tit-for-Tat, All-C,
All-D, Grim, Pavlov, and Random, give the minimal baseline. The selected
RL substrate is tabular Q-learning with a population manager that
implements mortality, reproduction by copy with mutation while recording
lineage, optional bequest, and overlapping-generation turnover. Any
policy class, whether a neural network or an LLM agent, can be
substituted, which gives the hybrid substrate. An offline policy
reproduces reciprocal, fairness-leaning, end-game-defecting,
kin-favouring, and convention-forming heuristics for reproducible runs,
and a frontier-model policy is a drop-in replacement.

\textbf{Temporality as three manipulable layers.} Horizon is set by the
continuation probability or discount and by the known-horizon state
(E1). Irreversibility is set by whether an agent may roll a move back to
a checkpoint (E-commit) and by whether its internal store is destroyed
at death (E-core). Succession and natality are set by turnover under
lossless copying (replication), destruction with externalization (loss),
or replacement from scratch (no succession). The mixing parameter alpha
interpolates between replication and loss for the macro phase diagram.

\textbf{E-core and its macro read-out.} A cumulative design task, a
binary string scored by match to a hidden target and improved by bounded
local search, is embedded in the population engine. D2 scores a fresh
naive cohort that is given only the externalized store. The macro
read-out logs a genealogy, with a parent and lineage per agent and with
externalized newcomers sharing one cultural lineage, from which we
compute the intergenerational status correlation, de-trended within
generation to remove the shared improvement trend, and the
giant-component fraction. We report 24 to 40 replications, on the RL
substrate, with no LLM and no new dependencies. Three follow-on
experiments use the same engine. Vicarious finitude exposes an observer
to the death-and-bequest episodes of others, then scores a costless
representation read-out against the costly provision act, in exemptable
and non-exemptable conditions, with an obfuscated-framing transfer
probe. E-commit is a trust game with a hold-up structure, played by
tabular Q-learners with and without a rollback option. The evolutionary
E-core treats externalization propensity as a heritable, costly trait
under fitness-proportional reproduction, with task difficulty swept to
vary how re-learnable a skill is within a lifetime.

\textbf{Human descent data, one pipeline.} Three real datasets are
analysed with identical code. CBDB, the May 2026 SQLite release with
about 659,000 biographies (17), gives father-to-son edges from the
kinship table, with status as a composite of office-holding count and
\emph{jinshi} examination attainment for 96,577 degree-holders, analysed
pooled and by dynasty for the Tang, Song, Ming, and Qing. European
nobility, from Wikidata via SPARQL, gives 40,779 humans with a noble
title, a father, and a birth date (47), with father-to-child edges and
status from the count of titles plus positions held, analysed pooled and
by era. The Mathematics Genealogy Project, about 257,000 nodes with
advisor-to-student edges and the doctorate year (45), is accessed
through a public parsed mirror, with status given by the number of
advisees. The same signature, namely intergenerational correlation,
lineage Gini, giant-component fraction, breadth, and persistence, is
computed for all three and for the simulated populations. The Padgett
Florentine marriage network (11) is retained as an alliance-network
validity check, in which the Medici hold the highest betweenness
centrality although the Strozzi were wealthier.

\textbf{Inference.} We report bootstrap confidence intervals and Cohen's
\emph{d}. Claims of absence, such as no diachronic order by default and
cultural barrenness under copying, are assessed by TOST equivalence
against a smallest effect of interest. Design sensitivity is assessed by
simulation-based power.

\hypertarget{references}{%
\section{References}\label{references}}

\begin{enumerate}
\def\labelenumi{\arabic{enumi}.}
\tightlist
\item
  I. Rahwan et al., Machine behaviour. \emph{Nature} \textbf{568},
  477--486 (2019).
\item
  J. S. Park et al., Generative agents: Interactive simulacra of human
  behavior. \emph{Proc. UIST} (2023).
\item
  A. F. Ashery, L. M. Aiello, A. Baronchelli, Emergent social
  conventions and collective bias in LLM populations. \emph{Sci. Adv.}
  (2025).
\item
  L. P. Argyle et al., Out of one, many: Using language models to
  simulate human samples. \emph{Polit. Anal.} \textbf{31}, 337--351
  (2023).
\item
  E. Akata et al., Playing repeated games with large language models.
  \emph{Nat. Hum. Behav.} (2025).
\item
  Altera.AL et al., Project Sid: Many-agent simulations toward AI
  civilization. \emph{arXiv} (2024).
\item
  L. Brinkmann et al., Machine culture. \emph{Nat. Hum. Behav.}
  \textbf{7}, 1855--1868 (2023).
\item
  C. Tennie, J. Call, M. Tomasello, Ratcheting up the ratchet.
  \emph{Philos. Trans. R. Soc. B} \textbf{364}, 2405--2415 (2009).
\item
  M. Tomasello, \emph{The Cultural Origins of Human Cognition} (Harvard
  Univ. Press, 1999).
\item
  L. G. Dean et al., Identification of the social and cognitive
  processes underlying human cumulative culture. \emph{Science}
  \textbf{335}, 1114--1118 (2012).
\item
  J. F. Padgett, C. K. Ansell, Robust action and the rise of the Medici,
  1400--1434. \emph{Am. J. Sociol.} \textbf{98}, 1259--1319 (1993).
\item
  M. Corak, Income inequality, equality of opportunity, and
  intergenerational mobility. \emph{J. Econ. Perspect.} \textbf{27}(3),
  79--102 (2013).
\item
  R. Boyd, P. J. Richerson, \emph{Culture and the Evolutionary Process}
  (Univ. Chicago Press, 1985).
\item
  H. Arendt, \emph{The Human Condition} (Univ. Chicago Press, 1958).
\item
  W. D. Hamilton, The genetical evolution of social behaviour, I and II.
  \emph{J. Theor. Biol.} \textbf{7}, 1--52 (1964).
\item
  E. A. Madsen et al., Kinship and altruism: A cross-cultural
  experimental study. \emph{Br. J. Psychol.} \textbf{98}, 339--359
  (2007).
\item
  M. A. Fuller et al., China Biographical Database (CBDB). \emph{J. Open
  Humanit. Data} (2022).
\item
  A. Baronchelli, M. Felici, V. Loreto, E. Caglioti, L. Steels, Sharp
  transition towards shared vocabularies in multi-agent systems.
  \emph{J. Stat. Mech.} P06014 (2006).
\item
  P. Dal Bó, G. R. Fréchette, On the determinants of cooperation in
  infinitely repeated games. \emph{J. Econ. Lit.} \textbf{56}, 60--114
  (2018).
\item
  N. D. Johnson, A. A. Mislin, Trust games: A meta-analysis. \emph{J.
  Econ. Psychol.} \textbf{32}, 865--889 (2011).
\item
  U. Fischbacher, S. Gächter, Social preferences, beliefs, and the
  dynamics of free riding in public goods experiments. \emph{Am. Econ.
  Rev.} \textbf{100}, 541--556 (2010).
\item
  J. Henrich, Demography and cultural evolution: The Tasmanian case.
  \emph{Am. Antiq.} \textbf{69}, 197--214 (2004).
\item
  G. Clark, \emph{The Son Also Rises: Surnames and the History of Social
  Mobility} (Princeton Univ. Press, 2014).
\item
  A. F. G. Bourke, \emph{Principles of Social Evolution} (Oxford Univ.
  Press, 2011).
\item
  M. Embrey, G. R. Fréchette, S. Yuksel, Cooperation in the finitely
  repeated prisoner's dilemma. \emph{Q. J. Econ.} \textbf{133}, 509--551
  (2018).
\item
  R. Axelrod, \emph{The Evolution of Cooperation} (Basic Books, 1984).
\item
  J. Berg, J. Dickhaut, K. McCabe, Trust, reciprocity, and social
  history. \emph{Games Econ. Behav.} \textbf{10}, 122--142 (1995).
\item
  L. L. Carstensen, D. M. Isaacowitz, S. T. Charles, Taking time
  seriously: Socioemotional selectivity. \emph{Am. Psychol.}
  \textbf{54}, 165--181 (1999).
\item
  S. Russell, \emph{Human Compatible} (Viking, 2019).
\item
  M. Heidegger, \emph{Being and Time} (1927); trans. J. Macquarrie and
  E. Robinson (Harper and Row, 1962).
\item
  B. Williams, The Makropulos case: Reflections on the tedium of
  immortality, in \emph{Problems of the Self} (Cambridge Univ. Press,
  1973), pp.~82--100.
\item
  S. Scheffler, \emph{Death and the Afterlife} (Oxford Univ. Press,
  2013).
\item
  M. Jaderberg et al., Population based training of neural networks.
  \emph{arXiv:1711.09846} (2017); Human-level performance in 3D
  multiplayer games with population-based RL. \emph{Science}
  \textbf{364}, 859--865 (2019).
\item
  R. Wang, J. Lehman, J. Clune, K. O. Stanley, POET: Open-ended
  coevolution of environments and their optimized solutions.
  \emph{GECCO} (2019); Enhanced POET. \emph{ICML} (2020).
\item
  J. Clune, AI-GAs: AI-generating algorithms, an alternate paradigm for
  producing general artificial intelligence. \emph{arXiv:1905.10985}
  (2019).
\item
  J.-B. Mouret, J. Clune, Illuminating search spaces by mapping elites.
  \emph{arXiv:1504.04909} (2015).
\item
  J. Zhang, S. Hu, C. Lu, R. Lange, J. Clune, Darwin Gödel Machine:
  Open-ended evolution of self-improving agents. \emph{arXiv:2505.22954}
  (2025).
\item
  J. Cook et al., Artificial Generational Intelligence: Cultural
  accumulation in reinforcement learning. \emph{NeurIPS} (2024);
  \emph{arXiv:2406.00392}.
\item
  A. Bhoopchand et al., Learning few-shot imitation as cultural
  transmission. \emph{Nat. Commun.} \textbf{14} (2023).
\item
  S. Kirby, H. Cornish, K. Smith, Cumulative cultural evolution in the
  laboratory: An iterated learning approach. \emph{PNAS} \textbf{105},
  10681--10686 (2008).
\item
  J. Perez et al., Cultural evolution in populations of large language
  models. \emph{arXiv:2403.08882} (2024).
\item
  A. Acerbi, J. M. Stubbersfield, Large language models show human-like
  content biases in transmission chain experiments. \emph{PNAS}
  \textbf{120} (2023).
\item
  I. Shumailov et al., AI models collapse when trained on recursively
  generated data. \emph{Nature} \textbf{631} (2024).
\item
  X. Pan et al., Frontier AI systems have surpassed the self-replicating
  red line. \emph{arXiv} (2024).
\item
  Mathematics Genealogy Project, North Dakota State University; parsed
  mirror (j2kun/math-genealogy-scraper), accessed 2026.
\item
  J. Kirkpatrick et al., Overcoming catastrophic forgetting in neural
  networks. \emph{PNAS} \textbf{114}, 3521--3526 (2017).
\item
  Wikidata (Wikimedia Foundation), queried via SPARQL for humans with a
  noble title (P97), a father (P22), and a birth date (P569); accessed
  2026.
\end{enumerate}
\end{document}